# QUANTUM COMPUTATION TOWARD QUANTUM GRAVITY*

P. A. Zizzi
Dipartimento di Astronomia dell' Università di Padova
Vicolo dell' Osservatorio, 5
35122 Padova, Italy
zizzi@giunone.pd.astro.it

## Abstract

The aim of this paper is to enlight the emerging relevance of Quantum Information Theory in the field of Quantum Gravity. As it was suggested by J. A. Wheeler, information theory must play a relevant role in understanding the foundations of Quantum Mechanics (the "It from bit" proposal). Here we suggest that quantum information must play a relevant role in Quantum Gravity (the "It from qubit" proposal). The conjecture is that Quantum Gravity, the theory which will reconcile Quantum Mechanics with General Relativity, can be formulated in terms of quantum bits of information (qubits) stored in space at the Planck scale. This conjecture is based on the following arguments: a) The holographic principle, b) The loop quantum gravity approach and spin networks, c) Quantum geometry and black hole entropy. From the above arguments, as they stand in the literature, it follows that the edges of spin networks pierce the black hole horizon and excite curvature degrees of freedom on the surface. These excitations are micro-states of Chern-Simons theory and account for the black hole entropy which turns out to be a quarter of the area of the horizon, (in units of Planck area), in accordance with the holographic principle. Moreover, the states which dominate the counting correspond to punctures of spin $j=1/2$ and one can in fact visualize each pixel of area as a bit of information. The obvious generalization of this result is to consider open spin networks with edges labeled by the spin -1/2 representation of SU(2) in a superposed state of spin "on" and spin "down". The pixel of area corresponding to such a puncture, will be "on" and "off" at the same time, and it will encode a qubit of information. This picture, when applied to quantum cosmology, describes an early inflationary universe which is a discrete version of the de Sitter universe.

## 1. INTRODUCTION

This is a short review paper based on my contributed talk to ICMP2000.
The aim of the talk was to enlight the relevance of Quantum Computation and Quantum Information in the context of Quantum Gravity. The talk was based on the results of three recent papers of mine: [1], [2], [3], where I investigated the three following topics, respectively: 1) Quantum Inflation, 2) Quantum Holography, and 3) The Quantum Gravity Register, which are also the contents of Sections 2, 3 and 4, respectively, in the present paper. Sect. 5 is devoted to some concluding remarks.
"Quantum Inflation" is a discrete version of the de Sitter Universe: Space and time are both discrete. Each time step is an integral multiple of the Planck time, and at each time step, the de Sitter horizon has a discrete area given in terms of pixels (a pixel is one unit of Planck area). The discrete entropy satisfies the holographic bound [4]. The cosmological constant is also quantized, and its outcoming present value is in accordance with inflationary theories.
"Quantum Holography" is the quantum version of the Holographic Principle [5], where instead of interpreting each pixel of area as a classical bit of information, one interprets it as a quantum bit (qubit). This is possible if the horizons' surfaces are pierced by edges of spin networks [6] [7] labeled by the spin-1/2 representation of SU(2) in the superposed state of spin "on" and spin "down". However, quantum holography is applicable only to the early inflationary universe, which was vacuum-dominated. At the end of inflation, the emergent environment caused decoherence and made the qubits collapse to classical bits.
A "Quantum Gravity Register" is a special kind of quantum memory register. It has two peculiar features: It is self-generating, and the evolution time is discrete. Thus, the usual quantum logic gates are replaced by discrete unitary evolution operators which connect Hilbert spaces of different dimensionality. A quantum gravity register, which grows with time, represents a early inflationary universe. The number of qubits wich decohered at the end of inflation, is responsible for the rather low entropy of our present universe.

## 2. QUANTUM INFLATION

### 2.1 Planckian foliation

In [1] we considered a 4-dimensional Riemannian space-time, with metric:

$$g_{\mu\nu}(\vec{x},t) = \begin{pmatrix} g_{00} = -1 & 0 \\ 0 & g_{ij}(\vec{x},t) \end{pmatrix} \qquad (i,j=1,2,3)$$

And performed a time-slicing of space-time.
The initial slice is at time $t_0 = t^*$ where $t^*$ is the Planck time:
$t^* \cong 5.3 \times 10^{-44}$ sec.

The lapse of time is $\Delta t = t^*$. The slice of order n is at time:

(1) $t_n = (n+1)t^*$ (n=0,1,2,…).

To each time step $t_n$ it corresponds a proper length: $L_n = ct_n = (n+1)L^*$, where $L^*$ is the Planck length: $L^* \cong 1.6 \times 10^{-33} cm$.

## 2.2 Quantum fluctuations

Let us consider the Wheeler relation : $\frac{\Delta g}{g} \cong \frac{L^*}{L}$ where $\frac{\Delta g}{g}$ is the quantum fluctuation of the metric and L is the linear extension of the region under study. If we take into account the above time slicing, the Wheeler relation becomes:

(2) $\left(\frac{\Delta g}{g}\right)_n = \frac{L^*}{L_n} = \frac{t^*}{t_n} = \frac{1}{n+1}$

For n=0, we recover: $\left(\frac{\Delta g}{g}\right)_0 = 1$, which means that at the Planck scale, the quantum fluctuation of the metric gets the maximum value.

## 2.3 The discrete energy spectrum

Following Wheeler, we believe that in quantum geometrodynamics, as well as in electrodynamics, when one examines a region of vacuum of dimension L, the fluctuation energy is of order:

$E \approx \frac{\hbar c}{L}$

Moreover, in quantum geometrodynamics there is a natural cut-off: the Planck length $L^*$.

In our model, we get the following discrete energy spectrum for the gravitational quantum fluctuations:

(3) $E_n \cong \frac{\hbar c}{L_n} = \frac{\hbar c}{(n+1)L^*} = \frac{E^*}{n+1}$

where $E^*$ is the Planck energy: $E^* \cong 1.2 \times 10^{19} GeV$. At the Planck scale (n=0), we recover $E_0 = E^*$.

## 2.4 The cosmological model

The resulting cosmological model is a discrete ensemble of de Sitter universes:

(4) $3H_n^2 = \Lambda_n c^2$

where $H_n \propto \frac{1}{(n+1)t^*}$ is the Hubble constant, and

(5) $\Lambda_n \propto \frac{1}{(n+1)^2 L^{*2}}$ is the quantized cosmological constant.

Today, the cosmological time is: $H^{-1} \cong 5 \times 10^{17} sec$, which corresponds to $n \cong 10^{60}$.

From eq. (5) it follows that the value of the cosmological constant is today:

$\Lambda_{NOW} \cong 5 \times 10^{-56} cm^{-2}$, in agreement with inflationary theories.

The area of a de Sitter horizon at time $t_n = (n+1)t^*$ is:

(6) $A_n = (n+1)^2 L^{*2}$, and the entropy satisfies the holographic bound [4]:

(7) $S_n = \frac{1}{4} A_n$.

## 3. QUANTUM HOLOGRAPHY

### 3.1 The Holographic Principle

The Holographic Priciple of 't Hooft and Susskind [5], is based on the Bekenstein bound [4]:

(8) $S = \frac{A}{4}$

where S is the entropy of a region of space of volume V, and A is the area, in Planck units, of the surface bounding V.

The entropy S of a quantum system is equal to the logaritm of the total number $N$ of degrees of freedom, i.e., the dimension of the Hilbert space:

(9) $S = \ln N = \ln(\dim H)$

In a discrete theory of N spins that can take only two values (Boolean variables) the dimension of the Hilbert space is $2^N$, hence the entropy directly counts the number of Boolean degrees of freedom:

(10) $S = N \ln 2$

From eqs. (8) and (10) one gets:

(11) $N = \frac{A}{4 \ln 2}$

Eq. (11) shows that, in a region of space-time surrounding a black hole, the number of Boolean degrees of freedom is proportional to the horizon area A.

't Hooft proposed that it must be possible to describe all phenomena within the bulk of a region of space of volume V by a set of degrees of freedom which reside on the boundary, and that this number should not be larger than one binary degree of freedom per Planck area.

All this can be interpreted as follows: each unit of Planck area (a pixel) is associated with a classical bit of information. The bit is the elementary quantity of information, which can take on one of two values, usually 0 and 1.

### 3.2 Spin networks

Spin networks are relevant for quantum geometry. They were invented by Penrose [6] in order to approach a drastic change in the concept of space-time, going from that of a smooth manifold to that of a discrete, purely combinatorial structure. Then, spin networks were re-discovered by Rovelli and Smolin [7] in the context of loop quantum gravity [8]. Basically, spin networks are graphs embedded in 3-space, with edges labelled by spins j=0, 1/2, 1, 3/2...and vertices labeled by intertwining operators. In loop quantum gravity, spin networks are eigenstates of the area and volume-operators [9].

If a single edge punctures a 2-surface transversely, it contributes an area proportional to:

(12) $\quad L^{*2}\sqrt{j(j+1)}$

where $L^*$ is the Planck length.
The points where the edges end on the surface are called "punctures".
If the surface is punctured in n points, the area is proportional to:

(13) $\quad L^{*2}\sum_{n}\sqrt{j_n(j_n+1)}$

Hence, gravity at the Planck scale is organized into branching flux tubes of *area*.

### 3.3 Spin networks and black hole entropy

What happens when this picture is applied to a black hole horizon? The flux lines pierce the black hole horizon and excite curvature degrees of freedom on the surface [10]. These excitations are described by Chern-Simons theory and account for the black hole entropy.

The very important feature of Chern-Simons theory is that it is possible to "count" the number of states: for a large number of punctures, the dimension of the Hilbert space $H_P$ for a permissible set P of punctures $P = \{j_{P_1},...j_{P_n}\}$ goes as:

(14) $\quad \dim H_P \approx \prod_{j_p \in P}(2j_p+1)$

The entropy of the black hole will be then:

(15) $\quad S = \ln\sum_P \dim H_P = const\,\dfrac{A}{4L^{*2}\gamma}$

where A is the area of the horizon, $L^*$ is the Planck length and $\gamma$ is a parameter of the theory called the Immirzi parameter [11].

Then, the best realization of the holographic hypothesis seems to be a topological quantum field theory like Chern-Simons. It should be noted that the states which dominate the counting of degrees of freedom correspond to punctures labelled by $j = \dfrac{1}{2}$. This fact reminds us of Wheeler's picture "it from Bit" [12] of the origin of black hole entropy.

A pixel can be either "*on*" $\equiv 1$ or "*off*" $\equiv 0$, where we take the convention that the pixel is "*on*" when the puncture is made by the edge of a spin network, ( in the spin-$\dfrac{1}{2}$ representation of SU(2)) in the state $\left|+\dfrac{1}{2}\right\rangle$ and that the pixel is "*off*" when the spin network's edge is in the state $\left|-\dfrac{1}{2}\right\rangle$.

### 3.4 Pixels as qubits

While the unit of classical information is the bit, endowed with Boolean logic, the unit of quantum information is the qubit, endowed with quantum logic.
A qubit differs from the classical bit in so far as it can be in both states $|0\rangle$ and $|1\rangle$ at the same time.

The most general 1-qubit is the superposed state: $a|0\rangle + b|1\rangle$ with the condition: $|a|^2 + |b|^2 = 1$, where a and b are the complex amplitudes of the two basis states.

A pixel can be "$on$"=1 and "$off$"=0 at the same time, [2] (i.e., it can be interpreted as a qubit), if the puncture is made by a (open) spin nework's edge in the superposed quantum state:

$$\frac{1}{\sqrt{2}}\left(\left|\frac{1}{2}\right\rangle \pm \left|-\frac{1}{2}\right\rangle\right).$$

If there are N such punctures $p_1, p_2, ....p_N$, the N pixels will be associated with N qubits.

In particular, let us consider now the discrete de Sitter horizons introduced in Sect. 2.

At time $t_n = (n+1)t^*$, the $n^{th}$ horizon has a surface area of $(n+1)^2$ pixels, which can be associated with N qubits, with $N = (n+1)^2$.

For n=0 (at the Planck time $t_0 = t^*$), N=1, we have only one puncture $p_1$ giving rise to one pixel of area, associated with the 1-qubit state:

(16)  $|1\rangle = \frac{1}{\sqrt{2}}(|on\rangle \pm |off\rangle)$

The state $|1\rangle$ of 1 qubit in eq. (16) represents the horizon state of a Euclidean Planckian black hole [1].

The ket $|1\rangle$ acts as a creation operator [2]:

$|1\rangle|N\rangle = |N+1\rangle$

and the bra $\langle 1|$ acts as a annihilation operator:

$\langle 1\|N\rangle = |N-1\rangle$

$\langle 1\|0\rangle = 0$

where the state $|N\rangle$ of N qubits is:

(17)  $|N\rangle = \frac{1}{\sqrt{2}^N}|1\rangle^N$

and represents the state of the $n^{th}$ de Sitter horizon, whose area is of $N = (n+1)^2$ pixels.

The dimension of the Hilbert space $H_N$ of N qubits is $2^N$.

## 4.  THE QUANTUM GRAVITY REGISTER

### 4.1  The quantum memory register

A quantum memory register is a system built of qubits. We will consider a quantum register of n qubits.

The state of n qubits is the unit vector in the $2^n$-dimensional complex Hilbert space:

$C^2 \otimes C^2 \otimes ... \otimes C^2$    n times.

As a natural basis, we take the computational basis, consisting of $2^n$ vectors, which correspond to $2^n$ classical strings of length n:

$|0\rangle \otimes |0\rangle \otimes ... \otimes |0\rangle \equiv |00...0\rangle$

$|0\rangle \otimes |0\rangle \otimes ... \otimes |1\rangle \equiv |00...1\rangle$

.
.
.

$|1\rangle \otimes |1\rangle \otimes ... \otimes |1\rangle \equiv |11...1\rangle$

For example, for n=2 the computational basis is:

$|0\rangle \otimes |0\rangle \equiv |00\rangle$

$|0\rangle \otimes |1\rangle \equiv |01\rangle$

$|1\rangle \otimes |0\rangle \equiv |10\rangle$

$|1\rangle \otimes |1\rangle \equiv |11\rangle$

In general, we will denote one basis vector of the state of n qubits as:

$|i_1\rangle \otimes |i_2\rangle \otimes ... \otimes |i_n\rangle \equiv |i_1 i_2 ... i_n\rangle \equiv |i\rangle$

where $i_1, i_2, ..., i_n$ is the binary representation of the integer i, a number between 0 and $2^{n-1}$. In this way, the quantum memory register can encode integers.

The general state is a complex unit vector in the Hilbert space, which is a linear superposition of the basis states:

$$\sum_{i=0}^{2^n-1} c_i |i\rangle$$

where $c_i$ are the complex amplitudes of the basis states $|i\rangle$, with the condition:

$$\sum_i |c_i|^2 = 1$$

For example, the most general state for n=1 is:

$c_0 |0\rangle + c_1 |1\rangle$

with: $|c_0|^2 + |c_1|^2 = 1$.

The uniform superposition $\frac{1}{\sqrt{2}}(|0\rangle + |1\rangle)$ is the one we will consider in the following, for the n=1 qubit.

To perform computation with qubits, we have to use quantum logic gates. A quantum logic gate on n qubits is a $2^n \times 2^n$ unitary matrix U.

The unitary matrix U is the time evolution operator which allows to compute the function f from n qubits to n qubits:

$|i_1 i_2 ... i_n\rangle \rightarrow U|i_1 i_2 ... i_n\rangle = |f(i_1 i_2 ... i_n)\rangle$

The hamiltonian H which generates the time evolution according to Schrodinger equation, is the solution of the equation:

$U = \exp(-\frac{i}{\hbar} \int H dt)$      with $UU^+ = I$

## 4.2  Quantum gravity computation

In our case, the quantum memory register is rather peculiar: it grows with time, and the time evolution is discrete. In fact, at each time step, a Planckian black hole, (the n=1 qubit state $|1\rangle$ which acts as a creation operator), supplies the quantum register with extra qubits.

At time $t_0 = t^*$ the quantum gravity register will consist of 1 qubit:

$$(|1\rangle|0\rangle)^1 = |1\rangle$$

At time $t_1 = 2t^*$ the quantum gravity register will consist of 4 qubits:

$$(|1\rangle|1\rangle)^2 = |2\rangle|2\rangle = |4\rangle$$

At time $t_2 = 3t^*$ the quantum gravity register will consist of 9 qubits:

$$(|1\rangle|2\rangle)^3 = |3\rangle|3\rangle|3\rangle = |9\rangle$$

At time $t_3 = 4t^*$, the quantum gravity register will consist 16 qubits:

$$(|1\rangle|3\rangle)^4 = |4\rangle|4\rangle|4\rangle|4\rangle = |16\rangle$$

At time $t_4 = 5t^*$, the quantum gravity register will consist of 25 qubits:

$$(|1\rangle|4\rangle)^5 = |5\rangle|5\rangle|5\rangle|5\rangle|5\rangle = |25\rangle$$

and so on.

The states $|1\rangle$, $|2\rangle$, $|3\rangle \ldots |n\rangle \ldots$ are the uniform superpositions:

$$|1\rangle = \frac{1}{\sqrt{2}}(|on\rangle + |off\rangle)$$

$$|2\rangle = \frac{1}{2}(|on\ on\rangle + |on\ off\rangle + |off\ on\rangle + |off\ off\rangle)$$

$$|3\rangle = \frac{1}{2\sqrt{2}}(|on\ on\ on\rangle + |on\ ono\ ff\rangle + |on\ off\ on\rangle + |on\ off\ off\rangle + |off\ on\ on\rangle + |off\ on\ off\rangle +$$

$$+ |off\ off\ on\rangle + |off\ off\ off\rangle)$$

and so on.

The general state $|n\rangle$ is:

$$|n\rangle = \frac{1}{\sqrt{2}^n}|1\rangle^{\otimes n}$$

At time $t_n = (n+1)t^*$ the quantum gravity register will consist of $(n+1)^2$ qubits:

$$(|1\rangle|n\rangle)^{n+1} = |n+1\rangle^{n+1} = |(n+1)^2\rangle$$

We call $|N\rangle$ the state $|(n+1)^2\rangle$, with $N = (n+1)^2$.

Now, let us consider a de Sitter horizon $|\Psi(t_n)\rangle$ [2] at time $t_n = (n+1)t^*$, with a discrete area $A_n = (n+1)^2 L^{*2}$ of $N$ pixels.

By the quantum holographic principle, we associate N qubits to the $n^{th}$ de Sitter horizon:

$$|N\rangle \equiv |\Psi(t_n)\rangle.$$

Let us remember that $|1\rangle = Had|0\rangle$ where *Had* is the Hadamard gate (which is a very important gate for quantum algorithms):

$$Had = \frac{1}{\sqrt{2}}\begin{pmatrix} 1 & 1 \\ 1 & -1 \end{pmatrix}$$

and $|0\rangle$ is the vacuum state, which can be identified either with the basis state $|on\rangle$ or with the basis state $|off\rangle$.

In fact, let us represent the basis states $|on\rangle$ and $|off\rangle$ as the vectors $\begin{pmatrix} 1 \\ 0 \end{pmatrix}$ and $\begin{pmatrix} 0 \\ 1 \end{pmatrix}$ respectively.

The action of $Had$ on the vacuum state $|0\rangle \equiv |off\rangle$ is:

$$Had|0\rangle = \frac{1}{\sqrt{2}}\begin{pmatrix} 1 & 1 \\ 1 & -1 \end{pmatrix}\begin{pmatrix} 0 \\ 1 \end{pmatrix} = \frac{1}{\sqrt{2}}\left[\begin{pmatrix} 1 \\ 0 \end{pmatrix} - \begin{pmatrix} 0 \\ 1 \end{pmatrix}\right] = |1\rangle^A,$$ where "A" stands for "antisymmetric".

The action of $Had$ on the vacuum state $|0\rangle \equiv |on\rangle$ is:

$$Had|0\rangle = \frac{1}{\sqrt{2}}\begin{pmatrix} 1 & 1 \\ 1 & -1 \end{pmatrix}\begin{pmatrix} 1 \\ 0 \end{pmatrix} = \frac{1}{\sqrt{2}}\left[\begin{pmatrix} 1 \\ 0 \end{pmatrix} + \begin{pmatrix} 0 \\ 1 \end{pmatrix}\right] = |1\rangle^S,$$ where "S" stands for "symmetric".

Then, the state $|N\rangle = |(n+1)^2\rangle$ can be expressed as:

(18) $\quad |N\rangle = (Had|0\rangle\frac{1}{\sqrt{2}^n}|1\rangle^n)^{n+1} = (Had|0\rangle)^{(n+1)^2} = (Had|0\rangle)^N$

As time is discrete, there will be no continuos time evolution, therefore there will not be a physical Hamiltonian which generates the time evolution according to Schrodinger equation. In [2], we considered discrete unitary evolution operators $E_{nm}$ between two Hilbert spaces $H_n$ and $H_m$ associated with two causally related "events": $|\Psi_n\rangle$ and $|\Psi_m\rangle$.

These "events" are de Sitter horizon states at times $t_n$ and $t_m$ respectively, with the causal relation: $|\Psi_n\rangle \leq |\Psi_m\rangle$, for $t_n \leq t_m$.

The discrete evolution operators:

(19) $\quad E_{nm} = |1\rangle^{(m-n)(m+n+2)} : H_n \to H_m$.

are the logic quantum gates for the quantum gravity register.

Thus we have:

(20) $\quad E_{nm} = |1\rangle^{(m-n)(m+n+2)} \equiv (Had|0\rangle)^{(m-n)(m+n+2)}$,

and the discrete time evolution is:

(21) $\quad E_{0n}|0\rangle = (Had|0\rangle)^{n(n+2)}|0\rangle = |1\rangle^{n(n+2)}|0\rangle = |1\rangle^{n(n+1)}|1\rangle|0\rangle = |1\rangle^{(n+1)^2} = |\Psi_{fin}\rangle$.

### 4.3 Decoherence

The quantum gravity register represents the early inflationary universe which is vacuum-dominated. Obviously then, during inflation, the qubits of the quantum gravity register cannot undergo environmental decoherence.
However, we know that at the end of the inflationary epoch, the universe reheated by getting energy from the vacuum, and started to be radiation-dominated becoming a Friedmann universe. This phase transition should correspond to decoherence.

According to inflationary theories, the end of inflation took place at time $t \approx 10^{-34}$ sec., which corresponds to $n = 10^9$ ($N = 10^{18}$ qubits).

If decoherence occurred now, at time $t_{now} = 10^{60} t^*$, corresponding to $N = 10^{120}$ qubits, the entropy of the universe would be, at present: $S_{MAX} = N \ln 2 \approx 10^{120}$.

Instead, we know [13] that the value of the entropy of the universe is, at present: $S_{now} = 10^{101} \sim 10^{102}$. In fact, it results:

(22)    $S_{MAX} / 10^{18} = S_{now} = 10^{102}$.

From the discrete energy spectrum in eq.(3), we get, for $n = 10^9$:

$E_{n=10^9} \approx 10^{11} GeV$, which is the reheating temperature at the end of inflation.

## 5. CONCLUSIONS

In this paper, we have summarized some recent work on the issues of quantum inflation, quantum holography and quantum-computing gravity.

The aim was to enlight the role of quantum computing in quantum gravity.

Quantum holography, where the pixels are interpreted in terms of qubits, (instead of bits) is relevant to model the very early universe, during the inflationary era. In fact, the resulting quantum computation is much faster than classical computation, and can account of the exponential expansion.

Of course, the quantum gravity computer is not like a common quantum computer.

The first peculiarity is that the time evolution of the quantum gravity register is discrete, and the quantum gates are discrete unitary operators which relate Hilbert spaces of different dimensionality.

The second peculiarity is that the quantum gravity register is autopoietic (it self-produces), and grows with time (at each time step $t_n$ there is an increase of 2n+3 qubits).

Eventually, in this quantum-computing approach to quantum gravity, the rather low value of the entropy of our present universe naturally emerges from the decoherence of $N = 10^{18}$ qubits at the end of inflation.

### AKNOWLEDGMENTS

I thank the ICMP Fellowship Commettee for EU support, and the Department of Astronomy, University of Padova, for hospitality.

**REFERENCES**

[1] P. A. Zizzi, "Quantum Foam and de Sitter-like Universe", hep-th/9808180; Int. J. Theor. Phys. 38 (1999) 2333.
[2] P. A. Zizzi, "Holography, Quantum Geometry, and Quantum Information Theory", gr-qc/9907063; Entropy, 2 (2000) 39.
[3] P. A. Zizzi, "Emergent Consciousness: From the Early Universe to our Mind", gr-qc/0007006.
[4] J. D. Bekenstein, Phys. Rev. D7 (1973) 2333.
[5] G. 't Hooft, "Dimensional reduction in quantum gravity", gr-qc/9310026;
G. 't Hooft, "The holographic Principle", hep-th/0003004;
L. Susskind, "The world as a hologram", hep-th/9409089
J. Bekenstein, "Holographic bound from Second Law of Thermodynamics", hep-th/0003058.
[6] R. Penrose, "Theory of quantised directions", in Quantum theory and beyond, ed. T. Bastin, Cambridge University Press (1971) 875.
[7] C. Rovelli and L. Smolin, "Spin networks and quantum gravity", gr-qc/9505006; Phys. Rev. D52 (1995) 5743.
[8] A. Ashtekar, "New variables for classical and quantum gravity", Phys. Rev. Lett. 57 (1986) 2244.
C. Rovelli and L. Smolin, "Loop representation of quantum general relativity", Nucl. Phys. B133 (1990) 80.
For recent reviews see, C. Rovelli, "Loop quantum gravity", gr-qc/9710008;
C. Rovelli, "Notes for a brief history of quantum gravity", gr-qc/0006061.
[9] C. Rovelli and L. smolin, "Discreteness of area and volume in quantum gravity", Nucl. Phys. B442 (1995) 593.
[10] A. Ashtekar, J. Baez, A. Corichi, and K. Krasnov, "Quantum Geometry and Black hole Entropy", gr-qc/9710007; Phys. Rev. Lett. 80 (1998) 904;
A. Ashtekar, J. Baez and K. Krasnov, "Quantum Geometry and Black Holes", gr-qc/9804039;
A. Ashtekar, J. Baez and K. Krasnov, "Quantum Geometry of Isolated Horizons and Black Hole Entropy", gr-qc/0005126.
[11] G. Immirzi, "Quantum Gravity and Regge calculus", gr-qc/9701052; Nucl. Phys. (1997) 65.
[12] J. A. Wheeler, "It from Bit", in Sakharov Memorial Lectures on Physics, Vol. 2, ed. L. Keldysh and v. feinberg, Nova Science, New York, 1992.
[13] R. Penrose, "The Emperor's New Mind", Oxford University Press, 1989.

.